\documentclass[prc,showpacs,showkeys,superscriptaddress,nofootinbib,twocolumn,floatfix]{revtex4}
\usepackage{amsfonts}
\usepackage{graphicx,amsmath,amssymb}
\usepackage[usenames,dvipsnames]{color}

\def\blfootnote{\xdef\@thefnmark{}\@footnotetext}

\newcommand{\lsim}{\lesssim}
\newcommand{\gsim}{\gtrsim}

\begin{document}

\title{Thermal Relaxation of Charm in Hadronic Matter}

\author{Min~He}
\affiliation{Cyclotron Institute and Department of Physics \& Astronomy, Texas A\&M University, College Station, TX 77843, USA}

\author{Rainer~J.~Fries}
\affiliation{Cyclotron Institute and Department of Physics \& Astronomy, Texas A\&M University, College Station, TX 77843, USA}
\affiliation{RIKEN/BNL Research Center, Brookhaven National Laboratory,
Upton, NY 11973, USA}

\author{Ralf~Rapp}
\affiliation{Cyclotron Institute and Department of Physics \& Astronomy, Texas A\&M University, College Station, TX 77843, USA}

\date{\today}

\begin{abstract}
The thermal relaxation rate of open-charm ($D$) mesons in hot and
dense hadronic matter is calculated using empirical elastic
scattering amplitudes. $D$-meson interactions with thermal pions are
approximated by $D^*$ resonances, while scattering off other hadrons
($K$, $\eta$, $\rho$, $\omega$, $K^*$, $N$, $\Delta$) is evaluated
using vacuum scattering amplitudes as available in the literature
based on effective Lagrangians and constrained by realistic
spectroscopy. The thermal relaxation time of $D$-mesons in a hot
$\pi$ gas is found to be around 25-50\,fm/$c$ for temperatures
$T$=150-180\,MeV, which reduces to 10-25\,fm/$c$ in a
hadron-resonance gas. The latter values, argued to be conservative
estimates, imply significant modifications of $D$-meson spectra in
heavy-ion collisions. Close to the critical temperature ($T_c$), the
spatial diffusion coefficient ($D_s$) is surprisingly similar to
recent calculations for charm quarks in the Quark-Gluon Plasma using
non-perturbative $T$-matrix interactions. This suggests a possibly
continuous minimum structure of $D_s$ around $T_c$.
\end{abstract}

\maketitle

\section{Introduction}
\label{introduction}
After the initial discovery of a new state of matter in high-energy
nuclear collisions at the Relativistic Heavy Ion Collider (RHIC),
the focus is now shifting to quantifying the properties of what is
believed to constitute a strongly coupled Quark-Gluon Plasma (sQGP).
Basic quantities characterizing the medium are its thermal spectral
functions and transport properties. In heavy-ion collisions (HICs),
the former are most directly studied in the electromagnetic (vector)
channel via the thermal emission of lepton pairs
(cf.~Ref.~\cite{Rapp:2009yu} for a recent review). The latter,
however, are best studied using observables with small but
controlled deviations from thermal equilibrium. Thus, heavy quarks
(charm ($c$) and bottom ($b$)), whose thermal equilibration time is
expected to be of the order of the QGP lifetime in HICs, are a
promising tool to quantify flavor transport, and eventually deduce
general properties of the QGP as formed in these
reactions~\cite{Rapp:2009my}.

The large masses of $c$ and $b$ quarks ($m_{c,b}$) enable us to
assess the modifications of their momentum spectra in HICs via a
diffusion process in an evolving background medium as formulated,
e.g., within a Fokker-Planck equation~\cite{Svetitsky:1987gq}
(typical early temperatures of the medium produced at RHIC,
$T\simeq250$\,MeV~\cite{Adare:2008fqa}, are well below
$m_{c,b}\simeq 1.5, 4.5$\,GeV). A reliable determination of the
heavy-quark (HQ) transport coefficients in the QGP depends on
several components. First and foremost these are microscopic
calculations of the thermal HQ relaxation rate in the
QGP~\cite{Svetitsky:1987gq,vanHees:2004gq,Moore:2004tg,Mustafa:2004dr,
vanHees:2005wb,vanHees:2007me,Gubser:2006bz,CasalderreySolana:2006rq,
Peshier:2008bg,CaronHuot:2009uh,Riek:2010fk} (see
Ref.~\cite{Rapp:2009my} for a review). Second, the coefficients need
to be implemented into a realistic bulk medium evolution (see, e.g.,
Ref.~\cite{Gossiaux:2011ea} for a recent discussion). Third,
heavy-flavor (HF) interactions in evolution phases other than the
QGP have to be evaluated, i.e., in the so-called pre-equilibrium
phase as well as in the hadronic phase. The former is of a
relatively short duration, $\Delta\tau_{\rm pre} \lsim 1$\,fm/$c$,
and is sometimes mimicked by reducing the formation time of the QGP.
The duration of the hadronic phase is substantially longer,
$\Delta\tau_{\rm had}$$\simeq$\,5-10\,fm/$c$. Its relevance for HF
phenomenology is further augmented by the fact that the hadronic
medium inherits the full momentum anisotropy from the QGP, believed
to be close to the finally observed one. Thus, even a rather weak
coupling of HF hadrons to hadronic matter can lead to noticable
contributions to their elliptic flow. Furthermore, if the QGP
realizes a minimum in its viscosity-to-entropy-density ratio,
$\eta/s$, close to the (pseudo-) critical temperature,
$T_c\simeq170$\,MeV, a hadronic liquid close to $T_c$ should possess
similar properties. This is usually referred to as a ``quark-hadron
duality", as suggested, e.g., in calculations of thermal dilepton
emission rates~\cite{Rapp:2009yu}.

Charm diffusion in hadronic matter has received little attention to date
(see Ref.~\cite{Laine:2011is} for a very recent estimate using heavy-meson
chiral perturbation theory). Its potential relevance has been noted in
Ref.~\cite{Rapp:2009my} based on calculations of $D$-meson spectral
functions in nuclear matter using effective hadron
Lagrangians~\cite{Lutz:2005vx,Tolos:2007vh}, as well as in a hot pion
gas~\cite{Fuchs:2004fh}. In the present paper, we augment these works to
evaluate charm diffusion in hadronic matter. Since the latter features
many resonances at temperatures approaching $T_c$, we not only utilize
$D\pi$ and $DN$ interactions but also scattering amplitudes off excited
hadrons ($K$, $\eta$, $\rho$, $\omega$, $K^*$, $\Delta$), as constructed
in the literature using effective Lagrangians and constrained by
charm-resonance spectroscopy. In this sense we provide a lower estimate
of the diffusion coefficient, based on existing elastic $D$-hadron
amplitudes.

Our paper is organized as follows. In Sec.~\ref{sec_Dint} we
``reconstruct" microscopic models for $D\pi$ scattering in a hot
pion gas (via $s$-channel resonances; Sec.~\ref{ssec_Dpi}), for $D$
scattering off strange and vector mesons (Sec.~\ref{ssec_DM}), and
off baryons (Sec.~\ref{ssec_DB}), by parameterizing pertinent
scattering amplitudes. In Sec.~\ref{sec_trans} these are applied to
calculate thermal $D$-meson relaxation rates and diffusion
coefficients in hot hadronic matter at vanishing chemical potential,
first in a pion gas (Sec.~\ref{ssec_Api}) and then in a resonance
gas (Sec.~\ref{ssec_Ahg}), and finally including chemical potentials
as appropriate for heavy-ion collisions at RHIC
(Sec.~\ref{ssec_Arhic}). Conclusions are given in
Sec.~\ref{sec_concl}.

\section{$D$-Meson Scattering and Width in Hot Hadronic Matter}
\label{sec_Dint}
In this section we recapitulate basic elements of the $D$-hadron
scattering amplitudes and apply them to calculate pertinent $D$-meson
widths in hadronic matter. We only employ known amplitudes from the
literature for the most abundant hadrons in a hot medium and combine
them into a lower-limit estimate for the $D$-meson width (and $D$-meson
relaxation rate in Sec.~\ref{sec_trans}). We first focus on pion-gas
effects, followed by interactions with strange and vector mesons, as
well as hot nuclear matter.

\subsection{Hot Pion Gas}
\label{ssec_Dpi}
\begin{figure}[!t]
\includegraphics[width=1\columnwidth
]{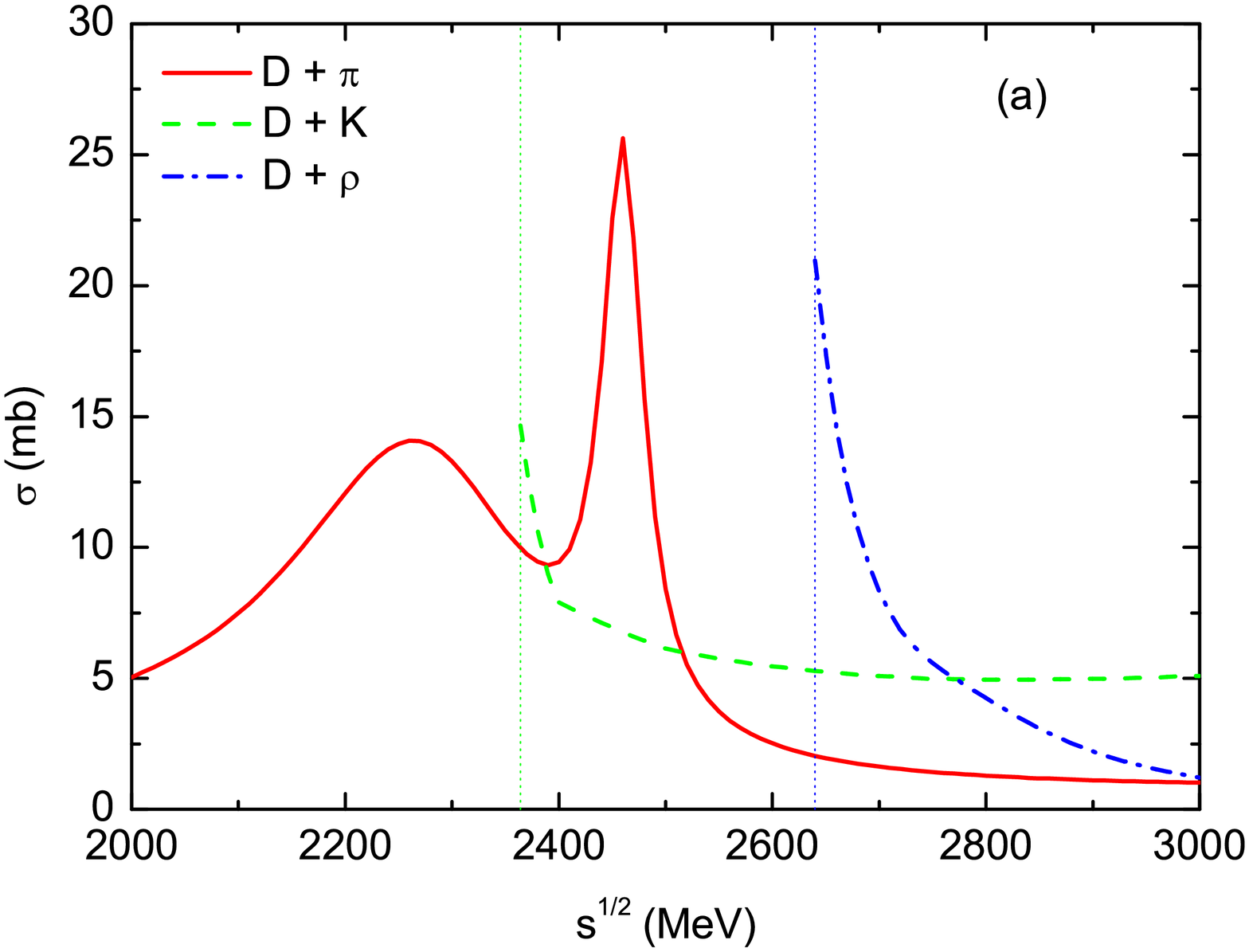}
\includegraphics[width=1\columnwidth
]{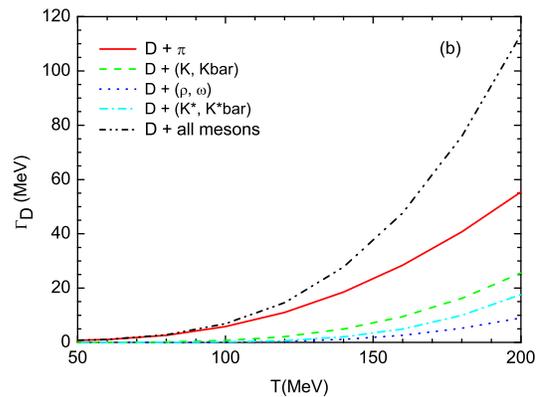}\caption{(a) Total elastic cross sections for
$D$-$\pi$ (solid line), $S$-wave $D$-$K$ (dashed line) and
$D$-$\rho$ (dash-dotted line) scattering based on the invariant
amplitudes as discussed in the text; vertical lines indicate the
respective thresholds. (b) Total $D$-meson scattering width in a
meson gas (dash double dotted line), consisting of contributions
from Goldstone bosons and vector mesons.} \label{fig_D-mes}
\end{figure}

Following Ref.~\cite{Fuchs:2004fh} $D$ interactions in the pion gas
are dominated by its chiral partner, $D_0^*(2308)$, by the vector
$D^*(2010)$ and by the tensor meson $D_2^*(2460)$.
Phenomenological control over their interaction vertices has become
possible due to new observations of $D$-meson resonances by the
BELLE Collaboration~\cite{Abe:2003zm}.
Especially, the large width of the $D_0^*(2308)$,
$\Gamma_0^{\rm tot}=276\pm99$\,MeV,
attributed to $S$-wave pion decay, leads to a large $D\pi D_0^*$
coupling constant. With the constituent-quark model assignment of
isospin $I$=1/2 for $D$-states, the pertinent forward scattering
amplitudes have been parameterized in Breit-Wigner form as
\begin{eqnarray}
\mathcal{M}(s,\Theta=0)=\sum_{j=0}^2 \frac{8\pi\sqrt{s}}{k}
\frac{(2j+1) \sqrt{s}\Gamma_j^{D\pi}}{s-M^2_j+i\sqrt{s}\Gamma_j^{\rm{tot}}}
\ , \label{MDpi}
\end{eqnarray}
where $\sqrt{s}$ and $k$ denote the center-of-mass energy and
3-momentum in the $D\pi$ collision, respectively, and
$j$ is the resonance spin.
The total resonance decay width $\Gamma_{\rm tot}$, is assumed to be
saturated by the partial one, $\Gamma_{0,1}^{D\pi}$, for $D^*$ and $D^*_0$,
while for the narrow state $D_2^*$ ($\Gamma_2^{\rm tot}=45.6\pm12.5$\,MeV)
a branching fraction estimated by the particle data group is
employed~\cite{pdg2010}. The resulting total $D\pi$ cross section is
displayed in the upper panel of Fig.~\ref{fig_D-mes}.

The $D$-meson self-energy in a pion gas can now be obtained by
the standard procedure of closing the in- and outgoing pion
lines of the forward $D\pi$ amplitude with a thermal
pion propagator. In the narrow-width approximation for the pion
propagator, the $D$-meson self-energy takes the form
\begin{equation}
\Pi(p_D;T)=\int\frac{d^3p_{\pi}}{(2\pi)^32E_{\pi}}f_B(E_{\pi};T)
\mathcal{M}(s,\Theta=0),
\label{Pi}
\end{equation}
where $f_B$ is the thermal Bose function and $E_{\pi}$ the on-shell pion
energy. Alternatively, one can obtain the collision rate (or on-shell
width), $\Gamma$=$-{\rm Im}\Pi(p_D^2$=$m_D^2,T)/m_D$ from the Boltzmann
equation as
\begin{eqnarray}
\Gamma(p_D;T)=\frac{\gamma_{\pi}}{2E_D(p_D)}\int\frac{1}{(2\pi)^9}
\frac{d^3q}{2E_\pi(q)} \frac{d^3p'}{2E_D(p')}
\frac{d^3q'}{2E_\pi(q')}
\nonumber\\
\times f_B(E_\pi(q);T) \ \overline{|\mathcal{M}|^2}\
(2\pi)^4\delta^4(p+q-p'-q'),
\label{Gamma2}
\end{eqnarray}
where $\gamma_{\pi}$=3 denotes the spin-isospin degeneracy factor of
the in-medium particle (pion), and $p,q$ and $p',q'$ are the momenta of
in- and outgoing particles, respectively. Equations~(\ref{Pi}) and
(\ref{Gamma2}) are related via the optical theorem (we have checked
their consistency). In the following we employ the latter since it is
close to the form of the thermal relaxation rate discussed in
Sec.~\ref{sec_trans} below.

The scattering width of a $D$-meson at rest in a pion gas is displayed
in the lower panel of Fig.~\ref{fig_D-mes}. For $T$=150-180\,MeV we find
$\Gamma_D$=20-40\,MeV,
where the chiral partner
of the $D$ provides the largest contribution through $S$-wave $D\pi$
scattering. For constant resonance widths, we find close agreement with
the results plotted in Fig.~2 of Ref.~\cite{Fuchs:2004fh}, while for
energy-momentum dependent widths (as quoted in Ref.~\cite{Fuchs:2004fh})
our results displayed in Fig.~\ref{fig_D-mes} turn out to be slightly
smaller (by ca.~20\%).

Chiral effective Lagrangians have been applied to $D$-meson
interactions with Goldstone bosons in
Refs.~\cite{Kolomeitsev:2003ac,Guo:2006fu,Gamermann:2006nm}. Once
the parameters are constrained by empirical information, the
resulting scattering amplitudes are very similar and closely agree
with the resonance ansatz of Ref.~\cite{Fuchs:2004fh} as employed
here. Chiral Lagrangians also predict non-resonant interactions. In
the (repulsive) $S$-wave $I$=3/2 $D\pi$ channel the scattering
length has been computed as $a_{D\pi}^{I=3/2}$$\equiv$$-f(s_{\rm
thr})$=$-$0.1\,fm in unitarized chiral perturbation theory
($\chi$PT)~\cite{Guo:2009ct} vs. $a_{D\pi}^{I=3/2}$=$-$0.15\,fm in
$\chi$PT to NNLO~\cite{Geng:2010vw}. We employ the former to
construct the $I$=3/2 scattering amplitude as
$\mathcal{M}(s)=8\pi\sqrt{s}f(s)$.


\subsection{Strange and Vector Meons}
\label{ssec_DM}
In a hot meson gas, the next abundant species after the pions
are the strangeness carrying Goldstone bosons and the light and
strange vector mesons.

For $D(K,\bar K)$ interactions we directly employ the results of the
(unitarized) chiral effective theory of
Ref.~\cite{Kolomeitsev:2003ac} (by parameterizing the amplitudes in
Figs.~1 and 2 in there). In the $S$-wave $DK$ channel, with
isospin-strangeness $(I,S)$=(0,1), the results are constrained by
the (loosely bound) $D_{s0}^*(2317)$ state (the analogue of
$D_0^*(2308)$ in $D\pi$), and again closely agree with
Refs.~\cite{Guo:2006fu,Gamermann:2006nm}. The application to the
$S$-wave $DK$ $(I,S)$=(1,1) and $D\bar K$ $(I,S)$=(0,$-$1) produces
Feshbach-type resonances (tetra-quarks) right at threshold
($E_{DK}^{\rm thr}$=2360\,MeV).
 For $D\bar K$ in the $(I,S)$=(1,$-$1) channel the Born amplitude is
predicted to be repulsive (analogous to $I$=3/2 $D\pi$); the
scattering length has been calculated as $a_{\rm D\bar
K}^{I=1}$=$-$0.22\,fm in unitarized $\chi$PT~\cite{Guo:2009ct} vs.
$a_{\rm D\bar K}^{I=1}$=$-$0.33\,fm in in $\chi$PT at
NNLO~\cite{Geng:2010vw}. Analogous to $I$=3/2 $D\pi$, we employ the
former to construct the pertinent scattering amplitude.

For $D\eta$ scattering we also adopt Ref.~\cite{Kolomeitsev:2003ac},
where the $S$-wave $(I,S)$=(1/2,0) amplitude is governed by a narrow
state of mass 2413\,MeV just below threshold.

The evaluation of $DV$ scattering ($V$=$\rho$, $\omega$, $K^*$)
requires to go beyond the chiral Lagrangian. This has been done in
Ref.~\cite{Gamermann:2007fi} starting from $SU(4)$ flavor symmetry
and then implementing chiral breaking terms. This framework,
properly unitarized, recovers the resonance poles computed with the
chiral Lagrangians, but extends to axialvector states coupling to
$S$-wave $DV$ interactions (in particular $D_1(2420)$, $D_1'(2427)$,
$D_{s1}(2460)$, $D_{s1}(2536)$). A convenient Breit-Wigner
parametrization of the elastic coupling of $DV$ to the dynamically
generated resonances has been quoted as
\begin{equation}
\mathcal{M}(s)=\frac{|g_{DV}|^2}{s-s_R} ,
\label{DV}
\end{equation}
where $g_{DV}$ is the dimensionful coupling constant and $s_R$ the
complex resonance-pole position. We include the three $I$=1/2
resonance couplings to $D\rho$ and $D\omega$ from Tab.~7 in
Ref.~\cite{Gamermann:2007fi}, two $I$=0 and two $I$=1 resonances
with $DK^*$ coupling (Tabs.~5 and 4 in \cite{Gamermann:2007fi},
respectively) and one $I$=0 state coupling to $D\bar{K}^*$ (Tab.~8
in \cite{Gamermann:2007fi}). As a representative, the isospin
$I$=1/2 $D\rho$ cross section is shown in the upper panel of
Fig.~\ref{fig_D-mes}.

The lower panel of Fig.~\ref{fig_D-mes} shows the temperature
dependence of mesonic contributions to the $D$-meson scattering
width as calculated from the above amplitudes. The width from
anti-/kaons is the next largest contribution after the pion. The
effect of vector mesons is smaller but significant, especially for
the $K^*$. The total $D$ width in a hot meson gas reaches
$\sim$80\,MeV around $T$$\simeq$180\,MeV, which should be a lower
limit since several channels have still been neglected, e.g., higher
partial waves (except for pions) and inelastic channels (e.g.
$DK^*\leftrightarrow D_s\pi$).

\subsection{Hot Nuclear Matter}
\label{ssec_DB}

To evaluate $D$ scattering off baryons we follow the same
strategy as for mesons, employing vacuum scattering amplitudes. More
elaborate many-body calculations for $D$-mesons in nuclear matter are
available~\cite{Lutz:2005vx,Tolos:2010rn}, but our procedure keeps
consistency with the mesonic sector and allows for an estimate of the
systematic error due to in-medium effects (e.g., selfconsistency
of selfenergy and in-medium scattering amplitudes).

\begin{figure}[!t]
\hspace{4mm}
\includegraphics[width=\columnwidth
]{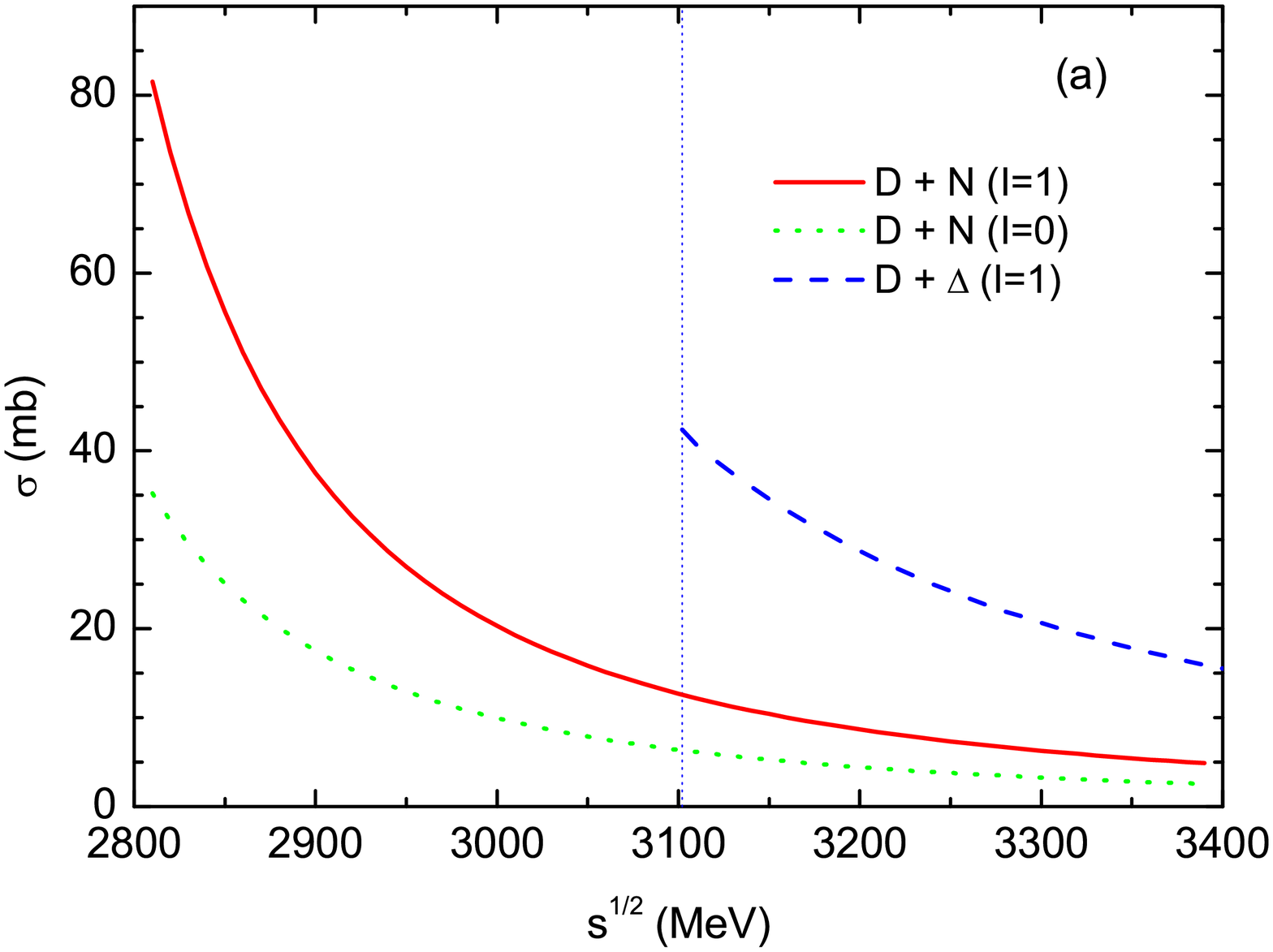}
\includegraphics[width=\columnwidth
]{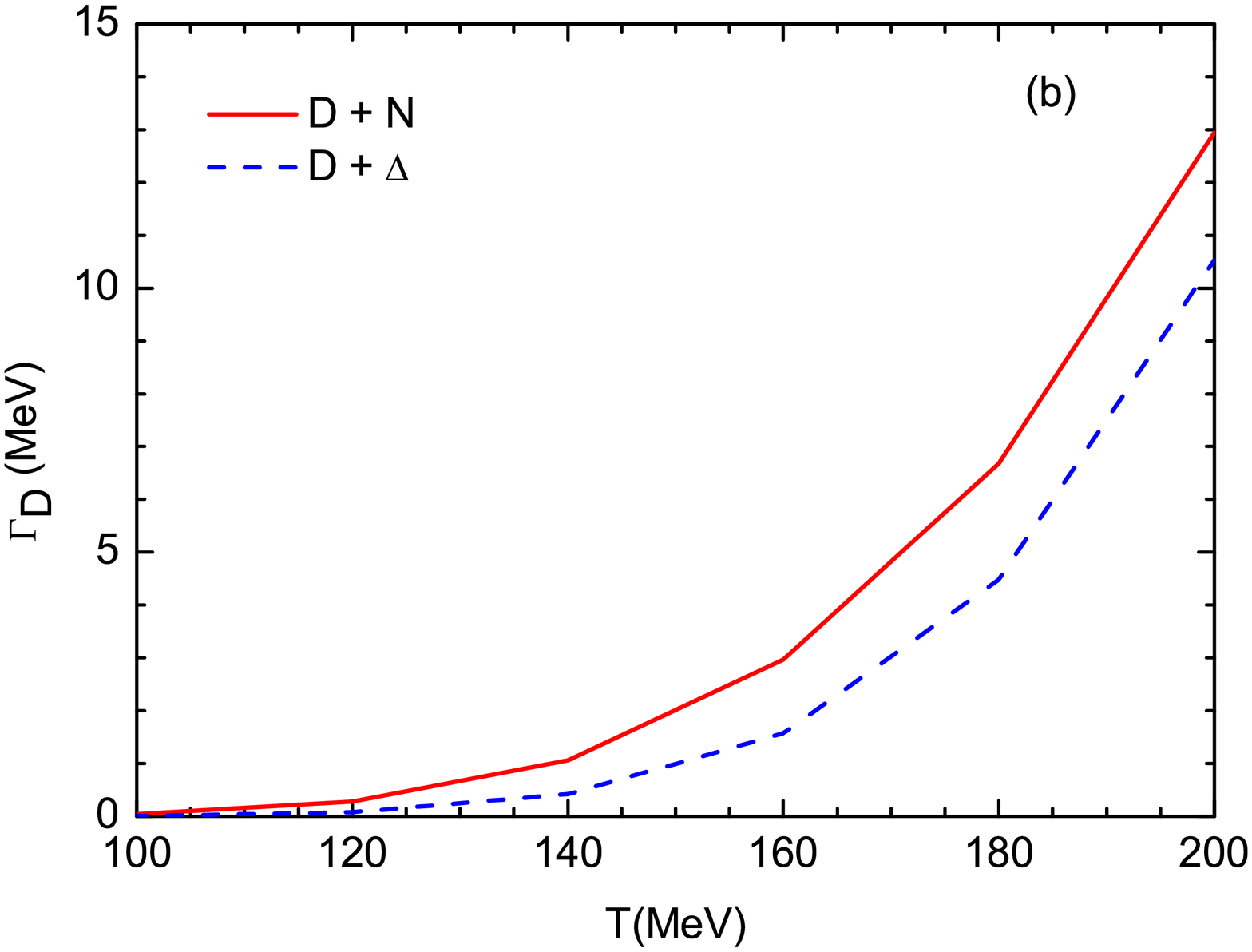} \caption{(a) Elastic cross sections for
$S$-wave $D$-$N$ scattering in the $I$=1 (solid line) and $I$=0
(dotted line) channel, and $D$-$\Delta$ scattering (dashed line,
$I$=1; vertical line: $D\Delta$ threshold) based on the invariant
amplitudes as discussed in the text; (b) $D$-meson widths from
$D$-$N$ scattering (solid line, including both $I=1$ and $I=0$
channels) and from $D$-$\Delta$ scattering (dashed line) at
vanishing chemical potential.} \label{fig_DB}
\end{figure}

We start from Ref.~\cite{Lutz:2005vx} where the $T$-matrix results of an
effective $DN$ interaction with coupled channels have been parameterized
in analogy to Eq.~(\ref{DV}) as
\begin{equation}
T(\sqrt{s})=\frac{|g|^2}{\sqrt{s}-m_R+ i\Gamma_R/2} \ ,
\label{DB}
\end{equation}
where $m_R$ is the resonance mass and $\Gamma_R$ the width. The
$T$-matrix, Eq.~(\ref{DB}), is related to the invariant scattering
amplitude through $\mathcal{M}(s)=N(s)T(s)$ with
$N(s)$=$(s+m_N^2-m_D^2+2m_N\sqrt{s})/(2\sqrt{s})$. The key states
are the dynamically generated $I$=0 $\Lambda_c(2595)$ and $I$=1
$\Sigma_c(2620)$ $S$-wave bound states. The former is experimentally
well established, while the latter is ca.~180\,MeV too deep compared
to the empirical $\Sigma_c(2800)$ state. However, the $I$=1 $DN$
cross section above threshold (shown in the upper panel of
Fig.~\ref{fig_DB}) is quite comparable to the results of a recent
meson-exchange model calculation~\cite{Haidenbauer:2010ch} in which
the $\Sigma_c$ is generated at its experimental mass (in fact, at
threshold the $I$=1 scattering length in
Ref.~\cite{Haidenbauer:2010ch} is significantly larger than for the
$\Sigma_c(2620)$ state of Ref.~\cite{Lutz:2005vx}). In the $I$=0
channel, the scattering lengths of Refs.~\cite{Lutz:2005vx} and
\cite{Haidenbauer:2010ch} agree well. The corresponding cross
section is also shown in the upper panel of Fig.~\ref{fig_DB}.

The $D\bar N$ scattering amplitude can be inferred from $\bar D N$ due
to $C$-symmetry of strong interactions. We have found no evidence in the
literature for (multi-quark) resonances in this system and adopt the
$D^-N$ elastic $S$-wave amplitude calculated in Ref.~\cite{Lutz:2005vx}.
The pertinent scattering and relaxation rates are about a factor of
$\sim$3 smaller than from $DN$ scattering.

The only available calculation of $D\Delta$ we are aware of has been
conducted in Ref.~\cite{Hofmann:2006qx}, within the same framework
as our $DN$ amplitudes are based on. In the only available $I$=1
$S$-wave channel the parametrization, Eq.~(\ref{DB}), reflects a
rather deep bound state ($m_R$=2613\,MeV, $\Gamma_R$$\simeq$0,
$g$=8.6). The cross section is shown in the upper panel of
Fig.~\ref{fig_DB}. Unlike the $D\bar N$ case, the $D\bar\Delta$
system is predicted to support a shallow $I$=1 bound state
($m_R$=2867\,MeV, $\Gamma_R$$\simeq$0, $g$=5.8). As a result, the
contribution of $D\bar\Delta$ scattering to the $D$-meson width and
relaxation rate is about half of that from $D\Delta$ scattering.

The lower panel of Fig.~\ref{fig_DB} shows that the width of a
$D$-meson at rest from scattering off thermally excited nucleons and
$\Delta$'s at vanishing chemical potential ($\mu_B$=0) is comparable
to that of light vector mesons (cf.~lower panel of
Fig.~\ref{fig_DB}). When adding $\bar N$ and $\bar\Delta$
contributions, the baryon-induced $D$-meson width computed here
amounts to ca.~15\,MeV at $T$$\simeq$180\,MeV. Note again that we
have neglected higher partial waves as well as higher excited
resonances including strange anti-/baryons.

It is instructive to compare our nucleon-induced width to a selfconsistent
many-body calculation~\cite{Tolos:2007vh}. From Fig.~6 in
Ref.~\cite{Tolos:2007vh}
we read off $\Gamma_D$$\simeq$100(80)\,MeV at $T$=100(150)\,MeV and
$\varrho_N$=$\varrho_0$, vs.~$\Gamma_D$$\simeq$75(65)\,MeV in our approach.
This indicates that neglecting in-medium effects does not lead to an
overestimate in our calculation.


\section{Thermal Relaxation in  Hadronic Matter}
\label{sec_trans}
The standard expression for the thermal relaxation rate of
a particle ($D$) in a heat bath in terms of its scattering
amplitude on medium particles ($h$) reads~\cite{Svetitsky:1987gq}
\begin{eqnarray}
A(p,T)&=&\frac{\gamma_{h}}{2E_D}\int\frac{1}{(2\pi)^9}
\frac{d^3q}{2E_h} \ \frac{d^3p'}{2E_D'} \ \frac{d^3q'}{2E_h'} f^h(E_h;T)
\nonumber\\
&& \hspace{-0.5cm}
\times \ \overline{|\mathcal{M}_{Dh}|^2}\
(2\pi)^4\delta^{(4)}(p+q-p'-q') (1-\frac{\vec p\cdot\vec p\,'}{\vec p^2})
\nonumber\\
&\equiv& \langle\langle 1-\frac{\vec p\cdot\vec p\,'}{\vec
p^2}\rangle\rangle \label{A}
\end{eqnarray}
with $\vec p$ ($\vec q$) and $\vec p\,'$ ($\vec q\,'$) being the
$D$-meson ($h$) momentum before and after the interaction,
respectively. The form of this expression is very similar to the one
for the total width, Eq.~(\ref{Gamma2}). The latter can be expressed
in terms of the average defined above as $\Gamma = \langle\langle 1
\rangle\rangle$. In Sec.~\ref{ssec_Api} we evaluate $A(p,T)$ for
$D$-mesons in a pion gas ($h$=$\pi$) and in Sec.~\ref{ssec_Ahg} for
all other hadrons whose amplitudes have been constructed in the
previous section.
In Sec.~\ref{ssec_Arhic} we evaluate the relaxation rate at
finite baryon and meson chemical potentials as applicable to the
hadronic phase below the chemical freeze-out temperature in heavy-ion
collisions at RHIC.

\subsection{Pion Gas}
\label{ssec_Api}
\begin{figure}[!t]
\hspace{4mm}
\includegraphics[width=1.1\columnwidth
]{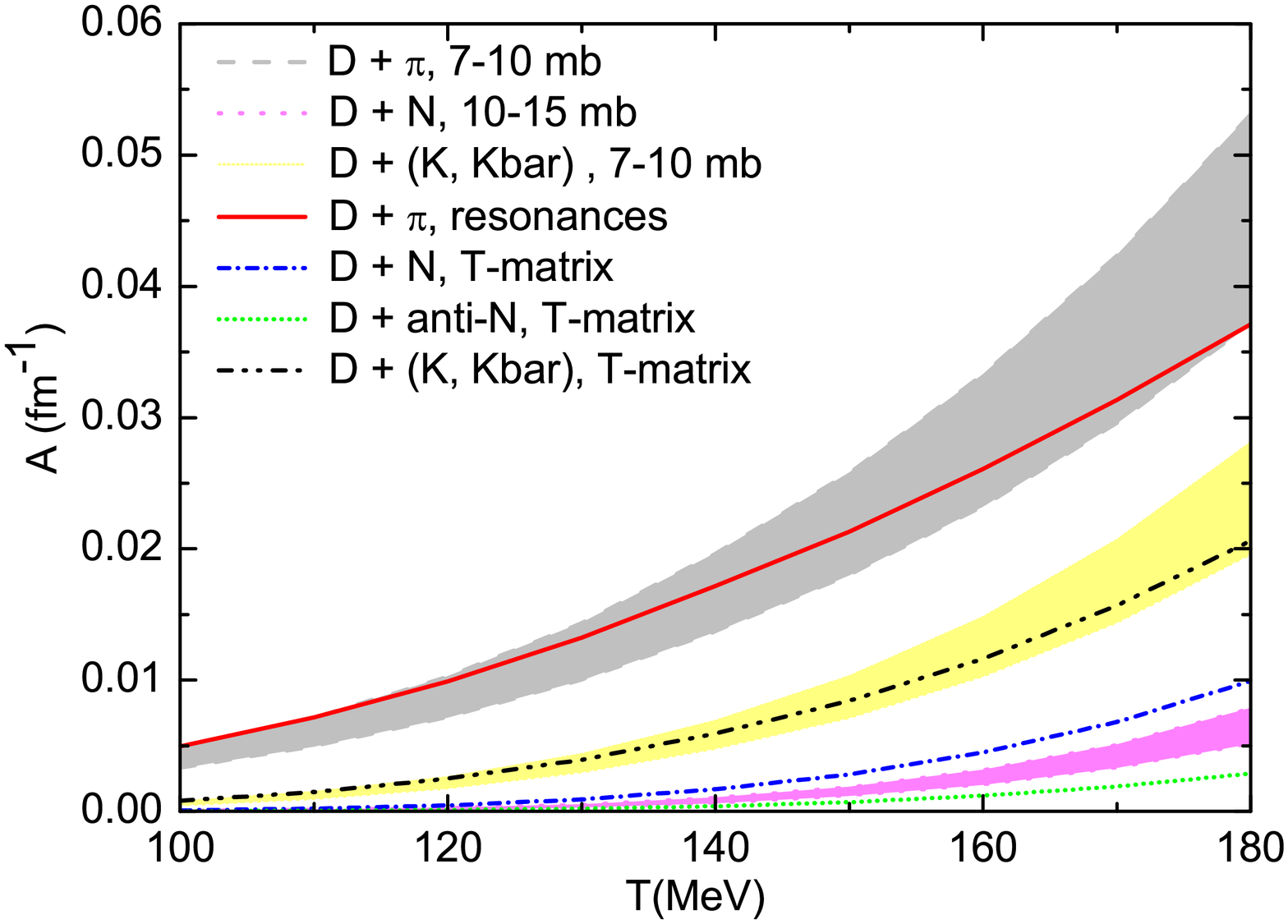} \caption{Thermal relaxation
rate of $D$-mesons at momentum $p_D$=0.1\,GeV in a gas of pions
(solid line), anti-/kaons (dash-double-dotted line), nucleons
(dash-dotted line) and antinucleons (dotted line), as a function of
temperature at vanishing chemical potentials using the elastic
scattering amplitudes constructed in Sec.~\ref{sec_Dint}. The bands
represent results obtained with constant $S$-wave $D\pi$, $DK$ and
$DN$ cross sections of 7-10\,mb and 10-15\,mb, respectively.}
\label{fig_ApiKN}
\end{figure}
The thermal relaxation rates for a $D$-meson at rest due to
scattering off pions in a thermal gas in chemical equilibrium
($\mu_\pi$=0), using the the scattering amplitude of
Eq.~(\ref{MDpi}), is displayed in Fig.~\ref{fig_ApiKN} as a function
of temperature. From $T$=100\,MeV to 180\,MeV the rate increases by
about a factor of 7, basically following the increase in pion
density from 0.2 to 1.4$\varrho_0$. Its magnitude at T=180\,MeV,
$A\simeq1/25$\,fm, is small but not negligible.
When replacing the $D\pi$ amplitude in Eq.~(\ref{Gamma2}) by one
yielding a constant $S$-wave cross section of
$\sigma_{D\pi}^S$=7-10\,mb, the pertinent band for the relaxation
rate essentially covers the result of our microscopic calculations.
The latter are closer to the upper end of the band at lower $T$ but
to the lower end at higher $T\gsim150$\,MeV. This reflects the
increased thermal motion of pions probing higher $\sqrt{s}$ in the
amplitudes where the latter decrease.

Compared to the recent work of Ref.~\cite{Laine:2011is}, where
$D$-meson diffusion in a lukewarm pion gas has been evaluated in
heavy-meson chiral perturbation theory, our value for the friction
coefficient is much smaller; e.g., at $T$=50(100)\,MeV,
Ref.~\cite{Laine:2011is} finds $A=\kappa/2m_DT\simeq 0.00055(0.05)$/fm,
about a factor of $\sim$4(10) larger than our pion-gas results,
$A\simeq0.00015(0.005)$/fm.

\subsection{Hadron Resonance Gas}
\label{ssec_Ahg}
\begin{figure}[!t]
\hspace{4mm}
\includegraphics[width=1.1\columnwidth
]{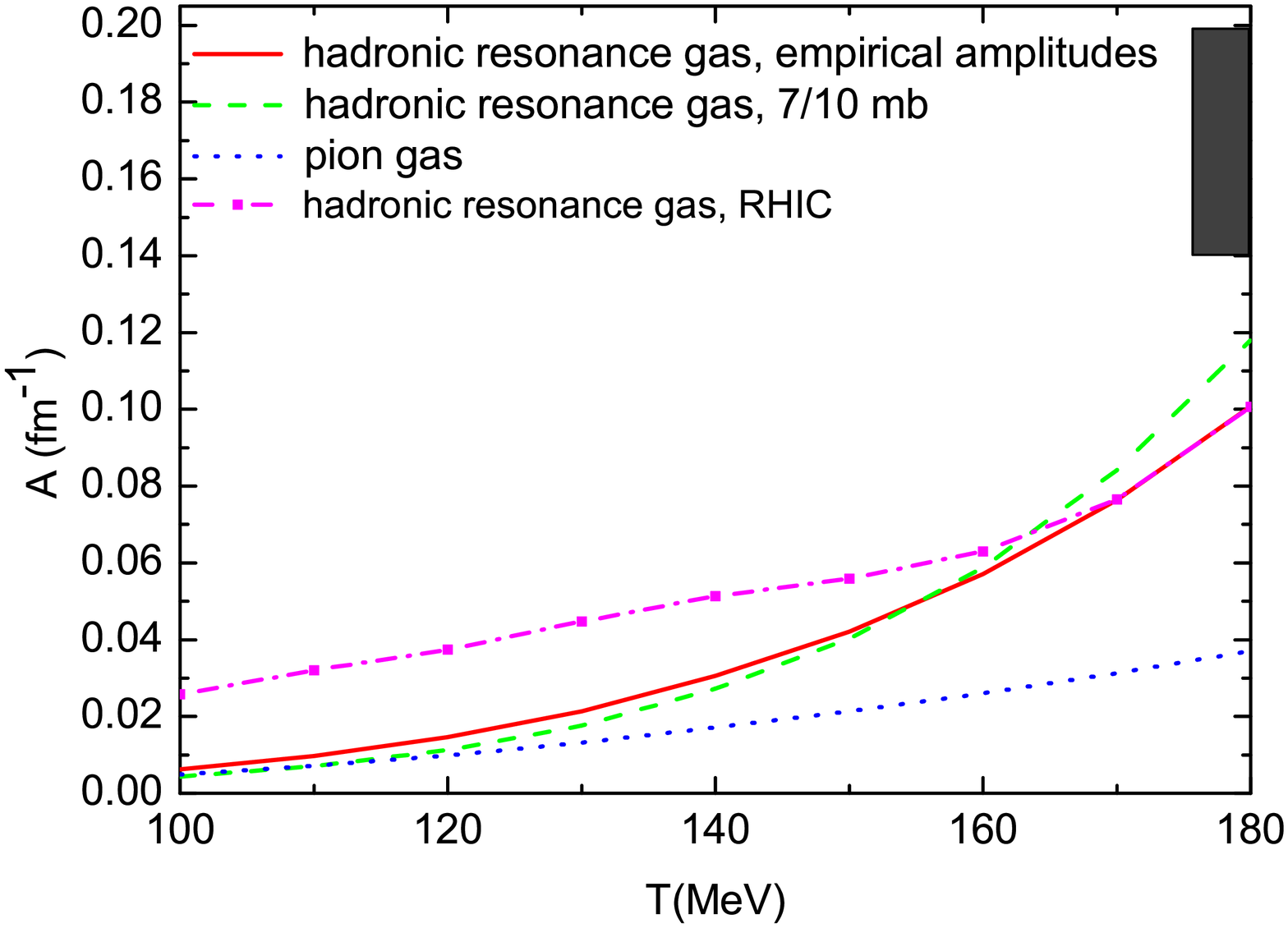}
\caption{Thermal relaxation rate of $p_D$=0.1\,GeV $D$-mesons using
empirical amplitudes in a hadron gas at vanishing (solid line) and
finite (dash-dotted line with squares) chemical potentials, as well as
in a pion gas (dotted line). The dashed line corresponds to isotropic
$D$-meson cross sections with mesons (7\,mb) and baryons (10\,mb).
The filled box at the upper right indicates
charm-quark relaxation rates in a QGP at 1.2\,$T_c$ from
in-medium $T$-matrix calculations~\cite{Riek:2010fk}.}
\label{fig_Ahrg}
\end{figure}
A hot hadron gas in equilibrium is characterized
by an increasing abundance of resonances with rising temperature.
For example, at $T$=180\,MeV, the density of baryons plus antibaryons
is above $\varrho_0$ and that of mesons with masses below 2\,GeV is
above 3$\varrho_0$. To improve the estimate of $D$-meson diffusion
in a pion gas for a more realistic hadron-resonance gas, we
include rescattering on all particles which at $T$=180\,MeV and
$\mu_h$=0 have a density at least 0.1$\varrho_0$, i.e., $\pi$, $K$,
$\eta$, $\rho$, $\omega$ and $K^*(892)$ in the meson sector and
anti-/nucleons, and $\Delta(1232)$, $\bar\Delta(1232)$ in the
anti-/baryon sector, using all scattering amplitudes of
Sec.~\ref{sec_Dint}.
The resulting $D$-meson friction coefficient in hadronic matter
at vanishing chemical potentials increases substantially over the
pion-gas result, by about a factor of $\sim$2(3) at $T$=150(180)\,MeV,
see Fig.~\ref{fig_Ahrg}. The individual contributions of $K+\bar K$
and $N$, $\bar N$ are compared to constant-cross-section calculations
in Fig.~\ref{fig_ApiKN}, indicating that a ``constituent" light-quark
cross section of 3-4\,mb
is compatible with our lower-limit estimates. A quantitative
decomposition of the individual hadron contributions to
kinetic $D$-meson relaxation at $T$=180\,MeV is given in Tab.~\ref{tab_A}.
Anti-/kaons provide the next-to-leading contributions after the pions,
while vector mesons, nucleons and Deltas play a smaller but non-negligible
role.
\begin{table}[!b]
\begin{tabular}{|c||c|c|}
  \hline
  Hadrons & $L_{I,2J}$ & $A$~[fm$^{-1}$] \\
  \hline\hline
  $\pi$ & $S_{1/2,0}$,~$P_{1/2,2}$,~$D_{1/2,4}$,~$S_{3/2,0}$ & $0.0371$\\
  \hline
  $K+\eta$ & $S_{0,0}$,~$S_{1,0}$ & $0.0236$\\
  \hline
  $\rho+\omega+K^*$ & $S_{1/2,2}$,~$S_{0,2}$,~$S_{1,2}$ & $0.0129$\\
  \hline
  $N+{\bar N}$ & $S_{0,1}$,~$S_{1,1}$ & $0.0128$\\
  \hline
  $\Delta+{\bar \Delta}$ & $S_{1,3}$ & $0.0144$\\
  \hline
\end{tabular}
\caption{Contributions to the thermal $D$-meson thermal relaxation
rate at $T$=180\,MeV indicating the quantum numbers of the included
scattering channels with $L$: partial wave, $I$: isospin and
$J$: total angular momentum.}
\label{tab_A}
\end{table}

As an estimate of medium effects in our vacuum $Dh$ amplitudes
we have performed a calculation for $A$ where we have introduced an
in-medium broadening of $\Gamma_R^{\rm med}$=200\,MeV into the Breit-Wigner
parameterizations. The final result for $A$ changes by less than 5\%.

Quantitatively, a value of $A$$\simeq$0.1/fm translates into a thermal
relaxation time of $\tau_D$$\simeq$10\,fm. This time scale is quite
comparable to lifetimes of the hadronic phase in Au-Au collisions at
RHIC. It is also similar to non-perturbative calculations of charm-quark
relaxation in the QGP using in-medium $T$-matrix
interactions~\cite{vanHees:2007me,Riek:2010fk}. This is encouraging
both from a conceptual (in the sense of a quark-hadron continuity
through $T_c$) and a practical point of view (relaxation rates of
this magnitude currently provide a fair phenomenology of current
heavy-flavor data at RHIC~\cite{vanHees:2005wb,vanHees:2007me}).

\begin{figure}[!t]
\hspace{4mm}
\includegraphics[width=1.1\columnwidth
]{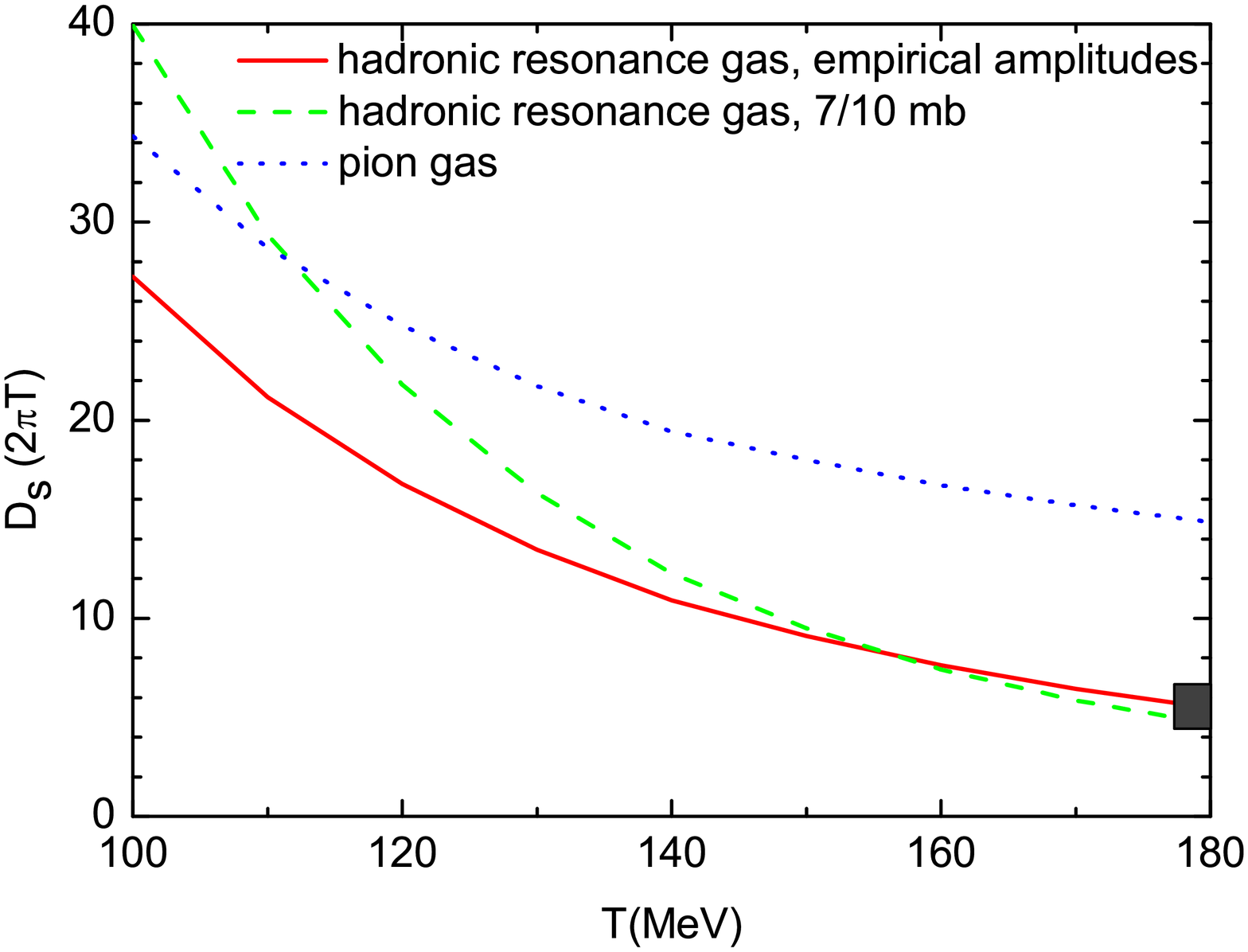}
\caption{Spatial $D$-meson diffusion coefficient
in units of the medium's thermal wavelength, $1/(2\pi T)$. The
filled box at the lower right indicates the range of values
calculated for charm quarks in the QGP at 1.2\,$T_c$ within an
in-medium $T$-matrix approach~\cite{Riek:2010fk}.}
\label{fig_Ds}
\end{figure}
In Fig.~\ref{fig_Ds} we display the spatial $D$-meson
diffusion coefficient, $D_s=T/(m_DA(p=0,T))$, in hadronic matter.
When normalized to the thermal wavelength, 1/(2$\pi T$), this
quantity decreases with $T$, reaching a value of $\sim$5 at
$T$=180\,MeV. Again, this is surprisingly close to $T$-matrix
results for charm quarks in the QGP~\cite{Riek:2010fk}, and,
together with those results, suggests a minimum across the
hadron-to-quark transition.

\subsection{RHIC Conditions}
\label{ssec_Arhic}
In relativistic HICs the chemical freeze-out
of hadron ratios~\cite{BraunMunzinger:2003zd} at a temperature of
$T_{\rm chem}$$\simeq$170\,MeV is significantly earlier than thermal
freeze-out of the light hadrons at $T_{\rm fo}$$\simeq$100\,MeV.
Therefore, to conserve the observed particle ratios in the hadronic
evolution, effective chemical potentials are required, reaching
appreciable values even at RHIC energies~\cite{Rapp:2002fc}, e.g.,
$\mu_\pi$($T$=100\,MeV)$\simeq$80\,MeV. We implement the chemical
potentials into the thermal hadron distribution functions and
recalculate the $D$-meson equilibration rate, Eq.~(\ref{A}). As a
result, the latter is enhanced at temperatures below $T_{\rm chem}$,
staying above 1/(25\,fm) for $T$$\ge$130\,MeV (cf.~Fig.~\ref{fig_Ahrg}),
implying noticeable modifications of $D$-meson spectra in the hadronic
phase of nuclear collisions at RHIC. For example, if the hadronic evolution
lasts for $\Delta\tau_{\rm had}$$\simeq$5\,fm, the expected modification
amounts to ca.~$(1-\exp[A~\Delta\tau_{\rm had}])\simeq20\%$.

\section{Summary and Conclusion}
\label{sec_concl}
We have evaluated kinetic transport of $D$-mesons in hot hadronic
matter by elastic scattering off the 10 most abundant hadron
species. The interaction strength with mesons and baryons has been
estimated from existing microscopic models for $D$-hadron
scattering, constrained by chiral symmetry and vacuum spectroscopy.
In a pion gas at $T$=100\,MeV, $D\pi$ resonance scattering leads to
a relaxation rate which is substantially smaller than what has been
found in a recent calculation using heavy-meson chiral perturbation
theory. Yet, when extrapolating our full results to temperatures in
the vicinity of $T_c$, the relaxation rate reaches ca.~0.1/fm,
translating into a spatial diffusion coefficient of
$D_s\simeq5/(2\pi T)$. This is comparable to non-perturbative
$T$-matrix calculations of charm-quark relaxation in the QGP. On the
one hand, this suggests a rather smooth evolution of charm transport
through $T_c$, i.e.,  a kind of ``duality" of hadronic and
quark-based calculations, reminiscent of what has been found for
dilepton emission rates. On the other hand, it implies that
quantitative calculations of $D$-meson spectra in heavy-ion
collisions have to account for hadronic diffusion. This insight is
reinforced once chemical freeze-out is implemented into the
evolution of the hadronic phase (via effective chemical potentials),
with an estimated modification of $D$-meson observables of at least
20\%. The apparent agreement of hadron- and quark-based approaches,
when extrapolated to around $T_c$, is encouraging, especially since
the magnitude of the transport coefficient is compatible with the
phenomenology of current heavy-flavor observables at RHIC. Our
findings thus pave the way for an improved theoretical accuracy
which will be needed to take advantage of upcoming precision
measurements at RHIC and LHC.

{\bf Note added.} Two subsequently submitted papers have also
addressed hadronic $D$-meson diffusion. In Ref.~\cite{Abreu:2011ic}
the use of unitarized chiral effective $D\pi$ interactions leads to
relaxation rates in a pion gas in close agreement with our results.
In Ref.~\cite{Ghosh:2011bw} $D$-hadron interactions have been
evaluated using Born amplitudes, leading to relaxation rates
significantly larger than our results.

\acknowledgments We thank L.~Tolos for informative discussions. This
work has been supported by U.S. National Science Foundation (NSF)
CAREER Award PHY-0847538, by NSF grant PHY-0969394, by the
A.-v.-Humboldt Foundation (Germany), by the RIKEN/BNL Research
Center and DOE grant DE-AC02-98CH10886, and the JET Collaboration
and DOE grant DE-FG02-10ER41682.

\end{document}